\documentclass[cmp]{svjour}

\usepackage{latexsym}
\usepackage{amsfonts}

\begin{document}

\title{$P(\phi )_2$ Quantum Field Theories and Segal's Axioms}

\author{Doug Pickrell}
\institute{Mathematics Department, University of Arizona, Tucson, AZ 85721, USA \\
\email{pickrell@math.arizona.edu}}

\maketitle

\begin{abstract} The purpose of this paper is to show that
$P(\phi )_2$ Euclidean quantum field theories satisfy axioms of
the type advocated by Graeme Segal.
\end{abstract}

\section{Introduction}
\label{sec0}

Throughout this paper, we fix (a bare mass) $m_0>0$, and a
polynomial $P:\mathbb R\to \mathbb R$ which is bounded from below.

If $\hat{\Sigma}$ is a closed Riemannian surface, the classical
$P(\phi )_2$-action is the local functional
\begin{equation}\mathcal A:\mathcal F(\hat{\Sigma })\to \mathbb R:\phi\to\int_{\hat{\Sigma}}
(\frac 12(\vert d\phi\vert^2+m_0^2\phi^2)+P(\phi
))dA,\label{0.1}\end{equation} where $\mathcal F(\hat{\Sigma })$
is the appropriate domain of $\mathbb R$-valued fields on
$\hat{\Sigma}$ for $\mathcal A$. A heuristic expression for the
$P(\phi )_2$-Feynmann-Kac measure is
\begin{equation}exp(-\mathcal A(\phi ))\prod_{x\in\hat{\Sigma}}d\lambda (\phi (x)),\label{0.2}\end{equation}
where $d\lambda (\phi (x))$ denotes Lebesgue measure for $\phi (x
)\in \mathbb R$.

It is notoriously difficult to understand the meaning of a generic
heuristic Feynmann-Kac expression.  Such an expression may not be
usefully represented by a measure at all.  However, for the
$P(\phi )_2$ action (\ref{0.1}), there is a well-known
interpretation of (\ref{0.2}), as a finite measure on generalized
functions,
\begin{equation}e^{-\int_{\hat{\Sigma}}:P(\phi ):_{C_0}}det_{\zeta}(m_0^2+\Delta
)^{-1/2}d\phi_C,\label{0.3}\end{equation} where $C_0=-\frac
1{2\pi}ln(m_0d(x,y))$, $C=(m_0^2+\Delta )^{-1}$, $d\phi_C$ is the
Gaussian probability measure with covariance $C$, $\int :P(\phi
):_{C_0}$ denotes a regularization of the nonlinear interaction,
and $det_{\zeta}$ denotes the zeta function determinant.

Our main purpose is to show that these Feynmann-Kac measures lead
naturally to a theory satisfying a primitive form of Segal's
axioms for a quantum field theory:  to a circle $S^1_R$ of radius
$R$, there is an associated Hilbert space, to a compact Riemannian
surface with geodesic boundary components there is an associated
operator, and these assignments have functorial properties
consistent with heuristic manipulations of path integrals.

The plan of the paper is the following.

In section \ref{sec1} we introduce some notation used throughout
the paper (we largely follow the conventions in \cite{GJ}).  We
also recall the primitive form of Segal's axioms, roughly
expressed above.

In section \ref{sec2}, and Appendix A, we discuss the $P(\phi
)_2$-Hilbert spaces.  The main point is that for $P(\phi )_2$
theories, in Segal's framework, the Hilbert space is independent
of $P$, $m_0$, and the metric on space (a union of circles).
Moreover, we can focus on the real part of the Hilbert space,
which simplifies matters somewhat.  This real Hilbert space is
defined in terms of the notion of the space of half-densities
associated to a measure class (Appendix A).

To define the vector that corresponds to a Riemannian surface with
geodesic boundary, in section \ref{sec3} we consider the
Feynmann-Kac measure which is associated to the double of the
surface (following \cite{GJ} or \cite{Simon}). The fundamental
result, established by constructive field theorists in the 70's,
is that (\ref{0.3}) is indeed a well-defined finite measure.

In section \ref{sec4}, we show that the Feynmann-Kac measures
naturally lead to a representation of Segal's category of compact
Riemannian surfaces with geodesic boundaries. The free case
($P=0$) has been considered previously, and more deeply, by Segal
(\cite{Segal1},\cite{Segal2}), and, from a different point of
view, by Dimock (\cite{Dimock}).  The main technical tool is the
work of Burghelea, Friedlander, and Kappeler on locality
properties of zeta function determinants (\cite{BFK}).

\section{Preliminaries}\label{sec1}

Throughout this paper all function spaces are real, and all
manifolds are oriented.

Suppose that $X$ is a closed Riemannian manifold.  The test
function space is $\mathcal D(X)=C^{\infty}(X;\mathbb R)$, with
the Frechet topology of uniform convergence of all derivatives. We
will write $f,g,h,..$ for test functions.  The space of
distributions is $\mathcal D'(X)$, with the weak topology relative
to $\mathcal D(X)$.  The Riemannian volume induces a map with
dense image
\begin{equation}\mathcal D(X)\to \mathcal D'(X):f\to fdV.\label{1.1}\end{equation}
We will write $\phi ,\psi ,..$ for distributions.  The pairing of
a test function and distribution will be denoted by $(f,\phi )$.

The positive Laplacian on functions will be denoted by $\Delta
=\Delta_X$, and $C(m,X)$ will denote the operator $(m^ 2+\Delta
)^{-d/2}$, where $d=dim(X)$.  In this paper we will only consider
$d=1,2$.  We will often abbreviate $C(m,X)$ to $C$, when there is
minimal risk of confusion.

The Gaussian probability measure on $\mathcal D'(X)$ with
Cameron-Martin Hilbert space
\begin{equation}W^{d/2}(X,m)=\{\phi :C(m,X)^{-1/2}\phi\in L^2(X,dV)\},\label{1.2a}\end{equation}
with inner product
\begin{equation}\langle\phi ,\psi\rangle_{W^{d/2}}=\int_XC^{-1/2}\phi C^{-1/2}\psi
dV,\label{1.2b}\end{equation} will be denoted by $d\phi_{C(m,X)}$.
Heuristically,
\begin{equation}d\phi_C=d\phi_{C(m,X)}=\frac 1{\mathcal Z}e^{-\frac 12\int_X\phi (m^
2+\Delta_X)^{d/2}\phi dV}d\lambda (\phi
),\label{1.3}\end{equation} where $d\lambda (\phi )$ denotes the
heuristic Riemannian volume on fields induced by $dV$; rigorously,
the Fourier transform is given by
\begin{equation}\int e^{-i(f,\phi )}d\phi_C=e^{-\frac 12(f,Cf)}.\label{1.4}\end{equation}

\begin{remark}\label{1.5}  (a) An $f\in \mathcal D(X)$ defines
a linear function $(f,\cdot )$ on $\mathcal D'(X)$.  One has
\begin{equation}\int\vert (f,\phi )\vert^2d\phi_C=\vert f\vert^2_{W^{-d/2}(X,m)}
.\label{1.6}\end{equation} Therefore there is an isometric
injection
\begin{equation}W^{-d/2}(X,m)\to L^2(d\phi_C)\label{1.7}\end{equation}
(and this can be extended to an isomorphism of Hilbert spaces
\begin{equation}\hat {S}(W^{-d/2}(X,m))\to L^2(d\phi_C),\label{1.8}\end{equation}
using normal ordering, where $\hat {S}(\cdot )$ denotes a Hilbert
space completion of the symmetric algebra).  Whereas we prefer to
parameterize the Gaussian $d\phi_C$ using the Cameron-Martin
Hilbert space $W^{d/2}(X,m)$, others prefer to think in terms of a
random process indexed by the dual Hilbert space $W^{-d/2}(X,m)$
(see chapter 1 of \cite{Simon} for a lucid discussion).

(b) Given $x\in X$, $\delta_x$ lies just outside of $W^{-d/2}$,
and hence does not quite define an $L^2$ random variable.  This is
one point of view on the main technical difficulty of quantum
field theory.
\end{remark}

\begin{lemma}\label{1.9}  If $\rho$ is a positive constant, and $\rho X$
denotes the space obtained by dilating all distances by $\rho$,
then
\begin{equation}d\phi_{C(m,\rho X)}=d\phi_{C(\rho m,X)}.\end{equation}
\end{lemma}

\begin{proof} Let $d=dim(X)$ and $dV_X$ the Riemannian volume
for $X$.  Then $dV_{\rho X}=\rho^ddV_X$, $\Delta_{\rho X}=
\rho^{-2}\Delta_X$, and the Cameron-Martin norm for
$d\phi_{C(m,\rho X)}$ equals
\begin{equation}\int_X\phi (m^2+\rho^{-2}\Delta_X)^{d/2}\phi\rho^ddV_X=\int\phi
((\rho m)^2+\vert\frac
{\partial}{\partial\theta}\vert^2)^{d/2}\phi
dV_X,\label{1.10}\end{equation} the Cameron-Martin norm for
$d\phi_{C(\rho m,X)}$.  \qed
\end{proof}

We will write $S^1_R$, rather than $RS^1$, to denote $S^1$ with
the metric $ds=Rd\theta$.

Suppose that $\Sigma$ is a compact Riemannian surface with
boundary, $S$.  We are assuming that $S$ has an intrinsic
orientation which, at a given point, may or may not agree with the
orientation induced by $\Sigma$.  We define $W^1(\Sigma ,m)$ to
consist of $L^2$ functions with locally $L^2$-integrable partial
derivatives such that the norm squared
\begin{equation}\int_{\Sigma}(d\phi\wedge *d\phi +*m^2\phi^2)=\int_{\Sigma}(\vert
d\phi\vert^2+m^2\phi^2)dA<\infty ,\label{1.11}\end{equation} where
$*=*_{\Sigma}$ denotes the star operator.  This is consistent with
(\ref{1.2a})-(\ref{1.2b}), when $S$ is empty.  As a topological
space, $W^1(\Sigma ,m)$ is independent of $m$.  When the specific
metric is not needed, we will simply write $W^1(\Sigma )$.

Because $S$ is smooth, smooth functions are dense in $W^1(\Sigma
)$.  The restriction map
\begin{equation}C^{\infty}(\Sigma )\to C^{\infty}(S)\label{1.12}\end{equation}
extends continuously to a map, the trace,
\begin{equation}W^1(\Sigma )\to W^{1/2}(S),\label{1.13}\end{equation}
The trace induces a short exact sequence of topological spaces,
\begin{equation}0\to W^1_0(\Sigma ,m)\to W^1(\Sigma ,m)\to W^{1/2}(S)\to 0.\label{1.14}\end{equation}
The orthogonal complement of the kernel is
\begin{equation}W^1_0(\Sigma ,m)^{\perp}=\{\phi\in W^1(\Sigma ):(m^2+\Delta )\phi
=0\quad in \quad\Sigma\setminus S\},\label{1.15}\end{equation} the
solution space of the Helmholtz equation. The quotient Hilbert
space structure on $W^{1/2}(S)$ is defined by a positive first
order pseudodifferential operator $D_{\Sigma}$ on $S$.  The
expression for this operator can be derived from the isomorphism
induced by the trace,
\begin{equation}W^1_0(\Sigma ,m)^{\perp}\to W^{1/2}(S):\Phi\to\phi =\Phi\vert_S
.\label{1.16}\end{equation} For a smooth solution $\Phi$ of the
Helmholtz equation, using Stokes's theorem,
\begin{equation}\int_{\Sigma}(d\Phi\wedge *_{\Sigma}d\Phi +m^2*_{\Sigma}\Phi^2)=\int_{\Sigma}
d(\Phi\wedge *_{\Sigma}d\Phi )=\int_{\partial\Sigma}\Phi\wedge *_{
\Sigma}d\Phi\label{1.17}\end{equation} (here $\partial\Sigma$
denotes the boundary with induced orientation). Consequently
\begin{equation}D_{\Sigma}\phi =\pm *_S(*_{\Sigma}d\Phi )\vert_S,\label{1.18}\end{equation}
where the sign is positive if the intrinsic and induced
orientations agree.  When $S$ is totally geodesic, this is simply
the unit outward normal derivative of $\Phi$ along $S$. The
operator $D_{\Sigma}$ is often referred to as the Dirichlet to
Neumann operator.  The principal symbol of the operator
$D_{\Sigma}^2$ is the induced metric on $T^{*}S$ (see subsection
$4.4$ of \cite{BFK}).

\subsection{Segal's definition (a primitive version)}

As in section 4 of \cite{Segal1}, let $\mathcal C_{metric}$ denote
the category for which the objects are oriented closed Riemannian
$1$-manifolds, and the morphisms are oriented compact Riemannian
$2$-manifolds with totally geodesic boundaries.

\begin{definition}\label{1.19}  A primitive $2$-dimensional
unitary quantum field theory is a representation of $\mathcal
C_{metric}$ by separable Hilbert spaces and Hilbert-Schmidt
operators such that disjoint union corresponds to tensor product,
orientation reversal corresponds to adjoint, $\mathcal
C_{metric}$-isomorphisms correspond to natural Hilbert space
isomorphisms.

\end{definition}

\begin{remark}\label{1.20} (a).  The naturality of the isomorphisms has
to be spelled out in terms of various commuting diagrams, which we
will leave to the reader's imagination (see section 4 of
\cite{Segal1} for some additional details).

(b) It is interesting to ask to what extent this definition
captures the notion of locality for a qft. Segal has recently
advocated additional axioms, which address the following two
(apparent) shortcomings: (1) a generic surface does not have many
closed geodesics, and in particular a morphism may not be
divisible (i.e. expressible as a composition);  and (2) a circle
can be cut into intervals, and the Hilbert space should be
recoverable from data associated to the intervals (see pages
424-425 of \cite{Segal1}).

(c) For a divisible morphism $\Sigma :S\to S$, the definition
implies that the corresponding operator is trace class. In this
case it also follows that the trace equals the partition function
of the closed surface obtained by sewing along $S$.

\end{remark}

To show that $P(\phi )_2$ satisfies this primitive form of Segal's
axioms, we will do the following.

To $S^1_R$ we will associate a real Hilbert space, which we will
ultimately denote by $\mathcal H(S^1)$, because this space will
not depend on $R$, $P$, or $m_0$.  This space will carry a natural
$Rot(S^1)$ action.  Since disjoint union of circles corresponds to
tensor product of Hilbert spaces, and a connected oriented
Riemannian $1$-manifold is isomorphic to $S^1_R$, for a uniquely
determined $R$, where the isomorphism is determined up to a
rotation, this determines the Hilbert space for more general
$1$-manifolds.  Since we will work with real Hilbert spaces, we
will not have to explicitly keep track of duals.

Let $\Sigma$ denote an oriented compact Riemannian surface with
geodesic and arclength parameterized boundary components.  A
component of $\partial\Sigma$ is said to be outgoing if the
parameterization agrees with the induced orientation, and ingoing
otherwise.  The union of outgoing boundary components will be
denoted by $(\partial\Sigma )_{out}$, and the union of ingoing
boundary components will be denote by $(\partial\Sigma )_{in}$. To
this surface we will associate a trace class operator
\begin{equation}\mathcal Z(\Sigma ):\mathcal H((\partial\Sigma )_{in})\to \mathcal H((\partial
\Sigma )_{out}).\label{1.21}\end{equation}

Let $\vert\Sigma\vert$ denote the morphism obtained from $\Sigma$
by reversing the orientation of all incoming circles. Because the
Hilbert spaces we consider are real, so that we can identify such
a space with its dual, there are equalities
\begin{equation}\mathcal Z(\Sigma )=\mathcal Z(\vert\Sigma\vert )\in \mathcal H(\partial\Sigma
)=\mathcal H(\partial\vert\Sigma\vert ).\label{1.22}\end{equation}

Suppose that $\Sigma_1$ and $\Sigma_2$ are two such surfaces, and
the number of outgoing boundary components of $\Sigma_1$ is the
same as the number of ingoing boundary components of $\Sigma_2$.
We can glue these Riemannian manifolds along
$(\partial\Sigma_1)_{out}$ and $(\partial\Sigma_2)_{in}$ to obtain
another such surface $\Sigma_2\circ\Sigma_1$.  We will show
\begin{equation}\mathcal Z(\Sigma_2\circ\Sigma_1)=\mathcal Z(\Sigma_2)\circ \mathcal Z(\Sigma_
1).\label{1.23}\end{equation}

\section{The Hilbert Space $\mathcal H(S^1)$.}\label{sec2}

To define the Hilbert space, we will use the notion of the space
of half-densities of a measure class.  This is described in
Appendix A.

Suppose that $M>0$.  For all of the $P(\phi )_2$ theories,
\begin{equation}\mathcal H(S^1_R)=\mathcal H(\mathcal C(M,S^1_R))\label{2.1}\end{equation}
where $\mathcal C(M,S^1_R)$ is the measure class on $\mathcal
D'(S^1)$ represented by the probability measure
$d\phi_{C(M,S^1_R)}$ on $\mathcal D'(S^1)$.

We also want to allow the possibility that $M=0$.  This is the
nonfinite measure
\begin{equation}d\phi_{C(0,S^1_R)}=\lim_{M\downarrow 0}\frac {\sqrt {2\pi}}Md\phi_{
C(M,R)}.\label{2.2}\end{equation}

A real generalized function on $S^1$ has a Fourier series
\begin{equation}\phi =\phi_0+\sum_1^{\infty}(\phi_ne^{in\theta}+\bar{\phi}_ne^{
-in\theta})\label{2.3}\end{equation} In these coordinates, if
$M>0$, $d\phi_{C(M,S^1_R)}$ is the infinite product of probability
measures
\begin{equation}d\phi_{C(M,S^1_R)}=\frac M{\sqrt {2\pi}}e^{-\frac 12M^2\phi_0^2}
d\lambda
(\phi_0)\prod_{n=1}^{\infty}d\mu_n^{(MR)}\label{2.4}\end{equation}
where
\begin{equation}d\mu_n^{(M)}=\frac {(M^2+n^2)^{1/2}}{2\pi}e^{-\frac 12(M^2+n^2)^{
1/2}\vert\phi_n\vert^2}d\lambda (\phi_n)\label{2.5}\end{equation}
If $M=0$, then
\begin{equation}d\phi_{C(0,S^1)}=d\lambda (\phi_0)\prod_{n=1}^{\infty}d\mu_n^{(
0)}.\label{2.6}\end{equation} Note there is no dependence on $R$
when $M=0$.

\begin{lemma}\label{2.7}The measure class $\mathcal C(M,S^1_R)$
is independent of $M\ge 0$ and $R$.
\end{lemma}

\begin{proof}In addressing this question, we can ignore the
$\phi_0$ factor.

Kakutani's theorem (Theorem 2.12.7, page 92, of \cite{B}), asserts
that the two infinite product measures
\begin{equation}\prod d\mu_n^{(mr)}\quad and\quad\prod d\mu_n^{(MR)}\label{2.8}\end{equation}
are either equivalent or disjoint, and they are equivalent if and
only if the inner product between the corresponding positive
half-densities is positive, i.e.
\begin{equation}\prod_{n=1}^{\infty}\int\sqrt {d\mu_n^{(mr)}d\mu_n^{(MR)}}>0.\label{2.9}\end{equation}
In doing this calculation, we can clearly assume $r=R=1$.

The $n^{th}$ factor of (\ref{2.9}) equals
\begin{equation}\frac {(M^2+n^2)^{1/4}(m^2+n^2)^{1/4}}{2\pi}\int_{\mathbb C}e^{-\frac
14((M^2+n^2)^{1/2}+(m^2+n^2)^{1/2})\vert x_n\vert^2}d\lambda (x_n
)\end{equation}
\begin{equation}=n(1+\frac {M^2}{n^2})^{1/2}(1+\frac {m^2}{n^2})^{1/2}\frac 2{(
M^2+n^2)^{1/2}+(m^2+n^2)^{1/2}}\label{2.11}\end{equation}
\begin{equation}=(1+\frac {M^2}{n^2})^{1/2}(1+\frac {m^2}{n^2})^{1/2}\frac 2{(1
+\frac {M^2}{n^2})^{1/2}+(1+\frac
{m^2}{n^2})^{1/2}}\label{2.12}\end{equation}

This has a positive infinite product over $n$.\qed
\end{proof}

We will need a more sophisticed result along these same lines.
Suppose that $D$ is a positive classical pseudodifferential
operator of order $1$ on $S$, a compact connected one-manifold
(e.g.  $D=(M^2+\Delta_{S^1})^{1/2}$). The principal symbol of the
operator $D^2$ determines a Riemannian metric on $S$, hence a
radius $R$.  By choosing an arclength coordinate $R\theta$, we can
suppose $S=S^1$ and the metric is $Rd\theta$.

\begin{proposition}\label{2.13} Let $D_1$ and $D_2$ denote two
operators as above such that $D_1$ and $D_2$ have the same
principal symbols.  Let $Rd\theta$ denote the corresponding
metric.  Then the Gaussian measures $\mu_i$ with Cameron-Martin
inner products
\begin{equation}\langle\phi ,\psi\rangle_i=\int_S\phi D_i\psi Rd\theta\end{equation}
are equivalent.
\end{proposition}

\begin{proof}Obviously
\begin{equation}\langle\phi ,\psi\rangle_2=\langle D^{-1}_1D_2\phi ,\psi\rangle_
1\label{2.14}\end{equation} Because $D_1$ and $D_2$ are classical
pseudodifferential operators, and they have the same principal
symbols,
\begin{equation}D_1^{-1}D_2=1+A\label{2.15}\end{equation}
where $A$ is a pseudodifferential operator of order $-1$. Because
$S$ is one dimensional, A is Hilbert-Schmidt. This implies that
the $\mu_i$ are equivalent (see Theorem 6.3.2, page 286, of
\cite{B}, or Theorem I.23, page 41, of \cite{Simon}).  \qed
\end{proof}

Since the Hilbert space corresponding to a circle is independent
of $R$, $M$, and $P$, we will denote it simply by $\mathcal
H(S^1)$.  More generally, given a closed $1$-manifold $S$, there
is a measure class associated to $W^{1/2}(S)$, and we will denote
the associated real Hilbert space of half-densities by $\mathcal
H(S)$.  This space is intrinsic to $S$, and it is naturally
isomorphic to the tensor product of the $\mathcal H(S_i)$, where
the $S_i$ (ordered in some way) denote the connected components of
$S$; see (5) of Appendix A.

\section{Feynmann-Kac Measures}\label{sec3}

To define the trace class operators corresponding to surfaces, we
will need a number of technical results about Feynmann-Kac
measures for closed Riemannian surfaces.

Suppose that $\hat{\Sigma}$ is a closed oriented Riemannian
surface. Let $\{f_k\}$ denote an orthonormal basis of real
eigenfunctions for the positive Laplace operator, $\Delta$, where
$\Delta f_k=\lambda_kf_k$, $0=\lambda_0<\lambda_1\le\lambda_ 2\le
..$.  A generalized function on $\hat{\Sigma}$ has an expansion
\begin{equation}\phi =\sum\phi_kf_k=\phi_0+\psi .\label{3.1}\end{equation}
In the coordinates $\phi_k\in \mathbb R$, $d\phi_{C(M,\hat{\Sigma
})}$ is the infinite product measure
\begin{equation}d\phi_{C(M,\hat{\Sigma })}=\prod d\mu^{(M)}_n\label{3.2}\end{equation}
where
\begin{equation}d\mu_n^{(M)}=\sqrt {\frac {M^2+\lambda_k}{2\pi}}e^{-\frac {(M^2
+\lambda_k)}2\phi_k^2}d\phi_k\label{3.3}\end{equation} We also
define
\begin{equation}d\phi_{C(0,\hat{\Sigma })}=d\lambda (\phi_0)\times d\psi_{C(0,\hat{
\Sigma })}=d\lambda (\phi_0)\prod_{k=1}^{\infty}\sqrt {\frac
{\lambda_ k}{2\pi}}e^{-\frac
{\lambda_k}2\phi_k^2}d\phi_k\label{3.4}\end{equation}

\begin{remark}\label{3.5} (a) The measure
$d\psi_{C(0,\hat{\Sigma } )}$ is a Gaussian measure on $\mathcal
D'(\hat{\Sigma })_0$, where $\psi\in \mathcal D'_ 0$ means $\psi
(1)=0$ $(\mathcal D_0'$ is the dual of $\mathcal D/\mathbb R$),
and the Cameron-Martin inner product is
\begin{equation}\int_{\hat{\Sigma}}d\psi_1\wedge *d\psi_2.\label{3.6}\end{equation}

(b) The space $\mathcal D'_0$ depends on the $C^{\infty}$
structure of $ \hat{\Sigma}$ (diffeomorphisms act naturally on
$\mathcal D/\mathbb R$, and hence its dual).  The Cameron-Martin
inner product depends on the conformal structure of $\hat{\Sigma}$
(because it involves the star operator on one-forms).  The
decomposition of distributions
\begin{equation}\mathcal D'(\hat{\Sigma })=\mathbb R\oplus \mathcal D'(\hat{\Sigma })_0:\phi
=\phi_0+\psi ,\label{3.7}\end{equation} as in (\ref{3.1}), depends
on the volume element of $\hat{\Sigma}$ (so that $\phi_0$ can be
interpreted as a distribution).  Consequently the measure
$d\phi_{C(0,\hat{\Sigma })}$ depends on the Riemannian structure
of $\hat{\Sigma}$; the measure $d\psi_{C(0,\hat{\Sigma } )}$ is
conformally invariant.
\end{remark}

Let $\mathcal C(M,\hat{\Sigma })$ denote the measure class of
$d\phi_{ C(M,\hat{\Sigma })}$.

\begin{lemma}\label{3.8} For constant $\rho >0$, the measure
class $\mathcal C(M,\rho\hat{\Sigma })$ is independent of $M\ge 0$
and $ \rho$.
\end{lemma}

\begin{proof} The independence of $M$ is essentially the same
as for Lemma \ref{2.7}.  The point is that $\lambda_k$ is
asymptotically $k$.

We again apply Kakutani's criterion for equivalence, as in
(\ref{2.9}).  We can also ignore the zero mode, $\phi_0$.

The $n^{th}$ factor, $\int\sqrt {d\mu_n^{(M)}d\mu_n^{(m)}}$ equals
\begin{equation}\frac {(M^2+\lambda_n)^{1/4}(m^2+\lambda_n)^{1/4}}{(2\pi )^{1/2}}
\int_{\mathbb R}e^{-\frac
14(M^2+\lambda_n+m^2+\lambda_n)\vert\phi_n \vert^2}d\lambda
(\phi_n)\label{3.9}\end{equation}
\begin{equation}=(\frac {\lambda_n}{2\pi})^{1/2}(1+\frac {M^2}{\lambda_n})^{1/4}
(1+\frac {m^2}{\lambda_n})^{1/4}(\frac {4\pi}{2\lambda_n+M^2+m^2}
)^{1/2}\label{3.10}\end{equation}
\begin{equation}=(1+\frac {M^2}{\lambda_n})^{1/4}(1+\frac {m^2}{\lambda_n})^{1/
4}(1+\frac {M^2+m^2}{2\lambda_n})^{-1/2}\label{3.11}\end{equation}
\begin{equation}=1+O(\frac 1{\lambda_n^2})\label{3.12}\end{equation}
Thus the inner product is positive, and the measures are
equivalent.

This proves the independence of $M$.  The independence of $\rho$
now follows from Lemma \ref{1.9}.  \qed
\end{proof}

\begin{remark}\label{3.13} If $\hat{\Sigma}$ is replaced by a
manifold of dimension $d$, and we consider an action defined by a
second order operator, then independence of mass holds if and only
if $d<4$, because $\lambda_n$ is asymptotic to $n^{2/ d}$.
\end{remark}

In the formulation of the following Lemma, we will use a basic
fact, due to Colella and Lanford, about the free field
$d\phi_{C(M,\hat{\Sigma })}$. This will be used frequently in the
remainder of the paper. A typical configuration $\phi$ for the
free field is not an ordinary function (or even a signed measure).
However, given a nice foliation of $\hat{\Sigma}$ by
$1$-submanifolds, a typical configuration can be thought of as a
continuous function (of a transverse parameter) with values in
distributions along the leaves.  A precise formulation of this, in
the case of $\mathbb R^2$, can be found in \cite{CL} (Theorem 1.1,
part (b), page 45, and see the paragraph following the Theorem,
for further comment).

\begin{lemma}\label{3.14}Suppose that $c:S^1_R\to\hat{\Sigma}$
is an isometric embedding.  Then the projection of
$d\phi_{C(M,\hat{\Sigma } )}$ to a measure on $\mathcal D'(S^1)$
belongs to the measure class $\mathcal C(M,S^1_R)$.
\end{lemma}

\begin{proof} Since $d\phi_{C(M,\hat{\Sigma })}$ is
a Gaussian measure, its projection must be a Gaussian measure. One
way to calculate the image of a Gaussian is to consider the map of
Cameron-Martin spaces, which in this case is the trace map
\begin{equation}W^1(\hat{\Sigma },M)\to W^{1/2}(S^1),\label{3.15}\end{equation}
where the inner product on the target is determined by a positive
first order pseudodifferential operator $D$, obtained by
considering the $W^1$ inner product on Helmholtz solutions on
$\hat{\Sigma}\setminus c(S^1)$, as in (\ref{1.14})-(\ref{1.18}).

To relate this directly to (\ref{1.14})-(\ref{1.18}), cut the
closed surface $\hat{\Sigma}$ along $c$ to obtain a compact
surface $\Sigma$ with two boundary components, one of which is
positively parameterized by $c$, and one of which is negatively
parameterized by $c$.  This induces a pseudodifferential operator
$D_{\Sigma}$ on $\partial\Sigma$, as in (\ref{1.14})-(\ref{1.18}).
This yields two pseudodifferential operators $D_{\pm}$ on $S^1$,
corresponding to the positive and negative $c$- parameterizations.
The operator $D=D_{+}+D_{-}$, by (\ref{1.18}).

Thus $D^2$ has principal symbol which is proportional to the
induced metric on $T^{*}S^1$, and the Lemma follows from
Proposition \ref{2.13} (recall also that the measure class
$\mathcal C(M,S^1_R)$ is independent of $M$ and $R$, by Lemma 2).

Alternatively, if $C(x,y)$ denotes the kernel for $C(M,\hat{\Sigma
} )$, the covariance for the projection is given by
\begin{equation}C'(\theta ,\theta')=C(c(\theta ),c(\theta')).\label{3.16}\end{equation}
One can read off the principal symbols for $C'$ and its inverse
$D$ from the fact that $C$ is asymptotically $-\frac
1{2\pi}ln(Md(x,y))$, as $d(x,y)\to 0$.  \qed
\end{proof}

\subsection{Normal Ordering}\label{normal ordering}

From now on we will use our fixed bare mass $m_0>0$. Let
$C=C(m_0,\hat{\Sigma })$.  If $f\in \mathcal D(\hat{\Sigma })$,
then $ (f,\cdot )\in L^2(d\phi_C)$.  By definition
\begin{equation}:(f,\cdot )^n:_C=H^{(f,Cf)}_n((f,\cdot ))\in L^2(d\phi_C),\label{3.17}\end{equation}
where $H_n^{\alpha}$ denotes the $n$th Hermite polynomial for the
Gaussian $(2\pi\alpha )^{-1/2}e^{-\frac 1{2\alpha}x^2}d\lambda (x
)$ (there are a number of different ways to motivate this
definition; see either section 6.3 of \cite{GJ} or chapter 1 of
\cite{Simon}). For example
\begin{equation}:(f,\cdot )^4:_C=(f,\cdot )^4-6(f,Cf)(f,\cdot )^2+3(f,Cf)^2\label{3.18}\end{equation}

One can define $:(f,\cdot )^n:_C$ equally well for $f\in
W^{-1}(\hat{ \Sigma })$, because of (\ref{1.7}).  Unfortunately,
given a point $x\in\hat{\Sigma}$, $ \delta_x$ is not in $W^{-1}$,
and in fact it is impossible to define $(\delta_x,\cdot )$ as a
random variable with respect to $d\phi_{C(m_0,\hat{\Sigma })}$
(the support of this measure consists of genuine distributions).
However, for $n\ge 0$, it is possible to define a regularization
$:(\delta_x,\cdot )^n:_C$, as a distribution; that is, given
$\rho\in \mathcal D(\hat{\Sigma })$,
\begin{equation}\int_{\hat{\Sigma}}:(\delta_x,\cdot )^n:_C\rho (x)dA(x)\label{3.20}\end{equation}
is a well-defined integrable random variable with respect to
$d\phi_C$.  For example (see section 8.5, page 152, of \cite{GJ}),
\begin{equation}:(\delta_x,\cdot )^4:_C=\lim_{t\downarrow 0}[(\delta_{t,x},\cdot
)^4-6(\delta_{t,x},C\delta_{t,x})(\delta_{t,x},\cdot
)^2+3(\delta_{ t,x},C\delta_{t,x})^2] \label{3.21}\end{equation}
where $\delta_{t,x}\in \mathcal D(\hat{\Sigma })$ satisfies
$\delta_{ t,x}\to\delta_x$ as $t\downarrow 0$. We will always
choose the functions $\delta_{t,x}$ to have compact support which
shrinks to $x$, and for these functions to depend smoothly on $x$.

Now suppose that we think of $C$ as a kernel function (which we
can do because we have a Riemannian background, and in particular
an area form).  A fundamental fact is that, near the diagonal,
\begin{equation}C(m_0,\hat{\Sigma })(x,y)=C_0(m_0,x,y)+C_f(m_0,x,y),\label{3.22}\end{equation}
where $C_0(m_0,x,y)=-\frac 1{2\pi}ln(m_0d(x,y))$ and $C_f$ is
smooth. We will often suppress the argument $m_0$.

For $\rho\in \mathcal D(\hat{\Sigma })$, we define
\begin{equation}\int :(\delta_x,\cdot )^n:_{C_0}\rho (x)dA(x)=\lim_{t\downarrow
0}\int H^{(\delta_{t,x},C_0\delta_{t,x})}_n((\delta_{t,x},\cdot )
)\rho (x)dA(x)\label{3.23}\end{equation} For example
\begin{equation}:(\delta_x,\cdot )^4:_{C_0}=\lim_{t\downarrow 0}[(\delta_{t,x},
\cdot )^4-6(\delta_{t,x},C_0\delta_{t,x})(\delta_{t,x},\cdot
)^2+3(\delta_{t,x},C_0\delta_{t,x})^2]\label{3.24}\end{equation}

\begin{remark}\label{3.25} This is local: the calculation of
$(\delta_{t,x},C_0\delta_{t,x})$ depends on arbitrarily small
neighborhoods of $x$ as $t\downarrow 0$. In a first version of
this paper, I claimed that one could just as well use $C$. But in
general this is false, because for fixed $x$, there is a constant
in the asymptotic expansion of $C$ (the value $C_f(x,x)$), which
is not zero, and which is not locally determined.
\end{remark}

One can also express (\ref{3.23}) in terms of regularization by
$C$: by a standard formula for `finite change of Wick order' (see
$(8.6.1)$ of \cite{GJ}), (\ref{3.23}) equals
\begin{equation}\sum_{j=0}^{[n/2]}\frac {n!}{(n-2j)!j!2^j}\int C_f(x,x)^j:(\delta_
x,\cdot )^{n-2j}:_C\rho (x)dA(x)\label{3.26}\end{equation} For
example
\begin{equation}:(\delta_x,\cdot )^4:_{C_0}=:(\delta_x,\cdot )^4:_C+6C_f(x,x):(
\delta_x,\cdot )^2:_C+3C_f(x,x)^2.\label{3.27}\end{equation} The
important point is that these regularizations agree up to lower
order terms.

In general we define $:P((\delta_x,\cdot )):_{C_0}$ by linear
extension. We will occasionally abbreviate this simply to
$:P:_{C_0}$, or, if we need to display the argument, to $:P(\phi
):_{C_0}$ [rather than the more cumbersome $:P((\delta_x,\cdot
)):_{C_0}(\phi )$].

The following is one of the fundamental results of constructive
quantum field theory.

\begin{theorem}\label{3.28}  Suppose that $P(\phi )$ is bounded below.
Then $exp(-\int_{\hat{\Sigma}}:P:_{C_0})\in L^1(d
\phi_{C(m_0,\hat{\Sigma })})$.
\end{theorem}

This follows, with relatively minor modifications, from the
arguments in section 8.6 of \cite{GJ}, or V.2 of \cite{Simon}
(Note that a closed Riemannian surface is conformally equivalent
to a constant curvature surface, and hence by uniformization can
be presented as a nice bounded region with generalized periodic
boundary conditions, and conformally Euclidean metric - with the
exception of the sphere).

\begin{definition}\label{3.29} The Feynmann-Kac measure for
$\hat{\Sigma}$ is the finite measure on $\mathcal D'(\hat{\Sigma }
)$
\begin{equation}e^{-\int_{\hat{\Sigma}}:P:_{C_0}dA(x)}det_{\zeta}(\Delta +m_0^2
)^{-1/2}d\phi_C.\end{equation}
\end{definition}

At a heuristic level, we can say that the $\zeta$-determinant is
essential because we have (for no good reason) normalized the free
background $d\phi_C$ to have unit mass; we have to add back in the
Gaussian volume of the Cameron-Martin space.

\section{Surfaces, Operators, and Sewing}\label{sec4}

Suppose that $\Sigma$ is a compact oriented Riemannian surface,
with geodesic and geodesically parameterized boundary components.
We also initially assume that all of the boundary components are
outgoing, i.e $\Sigma =\vert\Sigma\vert$.  We consider the closed
Riemannian surface
\begin{equation}\hat{\Sigma }=\Sigma^{*}\circ\Sigma ,\label{4.1}\end{equation}
where $\Sigma^{*}$ is the surface obtained by reversing the
orientation of everything.  Of fundamental importance is the
existence of a reflection symmetry through $\partial\Sigma$.

Let $S$ denote $\partial\Sigma$, and $C=C(m_0,\hat{\Sigma })$.  We
will write
\begin{equation}S_{*}(e^{-\int_{\hat{\Sigma}}:P:_{C_0}dA}d\phi_C)\label{4.2}\end{equation}
for the projection of this measure to a finite measure on
$\mathcal D'(S)$, which exists by Lemma \ref{3.14}.

\begin{definition}\label{4.3}For $\Sigma$ as above, we define
\begin{equation}\mathcal Z_1(\Sigma )=\mathcal Z_1(\vert\Sigma\vert )=(S_{*}(e^{-\int_{
\hat{\Sigma}}:P:_{C_0}dA}d\phi_C))^{1/2}\in \mathcal
H(S),\end{equation} and
\begin{equation}\mathcal Z(\Sigma )=det_{\zeta}(\Delta_{\hat{\Sigma}}+m_0^2)^{-1/4}
\mathcal Z_1(\Sigma )\in \mathcal H(S).\end{equation} For a closed
surface $\hat{\Sigma}$, we define $\mathcal Z(\hat{\Sigma } )$ to
be the integral of its Feynmann-Kac measure.
\end{definition}

Note that for a morphism $\Sigma :S_1\to S_2$, it follows
immediately from this definition that $\mathcal Z(\Sigma )$
represents a Hilbert-Schmidt operator.

\begin{theorem}\label{4.4} Suppose that $\Sigma_1$ and
$\Sigma_2$ are two morphisms which can be composed.  Then
\begin{equation}\mathcal Z(\Sigma_3)=\mathcal Z(\Sigma_2)\circ \mathcal Z(\Sigma_1),\end{equation}
where $\Sigma_3=\Sigma_2\circ\Sigma_1$.

(b) Suppose $\Sigma :S\to S$ is divisible. Then $\mathcal Z(\Sigma
)$ is trace class, and
\begin{equation}trace(\mathcal Z(\Sigma ))=\mathcal Z(\hat{\Sigma }),\end{equation}
where $\hat{\Sigma}$ is the closed surface obtained by gluing
$\Sigma$ to itself along $S$.
\end{theorem}

The rest of this section is devoted to the proof of this Theorem.
For (a) there are three possibilities: both of
$(\partial\Sigma_1)_{in}$ and $(\partial\Sigma_2)_{out}$ are
empty, one is empty, and neither is empty. The line of argument
for each of these cases is exactly the same, but the notational
details vary. We will carry out all the details for the second
possibility.

There are basically four parts to the argument.  In the first
part, we study the disintegration of the free Feynmann-Kac measure
with respect to its projection to a measure on generalized
functions on the boundary.  The second part involves the local
character of the nonlinear interaction. The third and fourth parts
are tightly intertwined:  these parts concern the sewing
properties for the normalized background Gaussian measures, and
the $\zeta$-regularized Gaussian volumes, respectively.

\subsection{Part 1.  Decomposition of free backgrounds
relative to traces}\label{subs4.1}

As above, we initially suppose that $\Sigma =\vert\Sigma\vert$,
with outgoing boundary $S$.  The trace map
\begin{equation}W^1(\hat{\Sigma })\to W^{1/2}(S):\hat{\phi}\to\hat{\phi}_S\label{4.5}\end{equation}
corresponds to a Hilbert space decomposition
\begin{equation}W^1(\hat{\Sigma },m_0)=W^1_0(\hat{\Sigma },m_0)\oplus W^1_0(\hat{
\Sigma },m_0)^{\perp}.\label{4.6}\end{equation} In turn,
\begin{equation}W^1_0(\hat{\Sigma },m_0)=W^1_0(\Sigma ,m_0)\oplus W^1_0(\Sigma^{
*},m_0)\label{4.7}\end{equation} and
\begin{equation}W^1_0(\hat{\Sigma },m_0)^{\perp}=W^1(\hat{\Sigma },m_0)\cap ker
(\Delta +m_0^2)\vert_{\hat{\Sigma}\setminus
S}.\label{4.8}\end{equation} The latter space has two other
realizations. On the one hand it is essentially isomorphic to
\begin{equation}W^1(\Sigma ,m_0)\cap ker(\Delta +m_0^2)\vert_{\Sigma\setminus S}
,\label{ 4.9}\end{equation} because a $W^1$-solution
$\hat{\phi}_s$ of the Helmholtz equation on $\hat{\Sigma}\setminus
S$ is necessarily even, i.e.  invariant with respect to the mirror
symmetry of $\hat{\Sigma}$ through $S$ (the even and odd parts of
a Helmholtz solution would also be solutions; the odd part
vanishes on $S$, hence it must be identically zero); hence
$\hat{\phi}_s$ is determined by its restriction to $ \Sigma$,
which we denote by $\phi_s$.  On the other hand it is also
isomorphic to $W^{1/2}(S)$, with the inner product determined by
$2D_{\Sigma}$, as in (\ref{1.18}).

We now want to apply these Hilbert space decompositions to obtain
decompositions of the corresponding Gaussian measures, in
particular our background Gaussian measures.  In the following we
will have to distinguish, for example, between $\hat{\phi}\in
W^1(\hat{ \Sigma },m_0)$, and a typical $\hat{\phi}$ in the
support of $d\hat{\phi}_C$; we will refer to the latter as a
random field (rather than introducing some additional notation).
We will also implicitly invoke the theorem of Collella-Lansford,
which, for example, allows us to make sense of the restriction of
a random $\hat{\phi}$ to $\Sigma$ or $S$.

The Gaussian measure $d\hat{\phi}_{C(m_0,\hat{\Sigma })}$ has a
disintegration relative to its projection to fields on $S$:
\begin{equation}d\hat{\phi}_{C(m_0,\hat{\Sigma })}=\int [d\hat{\phi}_{C(m_0,\hat{
\Sigma })}\vert\hat{\phi}_S=\phi_1]d(S_{*}(d\hat{\phi}_{C(m_
0,\hat{\Sigma })}))(\phi_1).\label{4.10}\end{equation} The
existence of this disintegration is a general fact (Proposition
13, section 2, No. 7, of \cite{Bo}). But as we will explain in the
following paragraphs, the `normalized conditional measure'
$[d\hat{\phi}_C\vert\hat{\phi}_ S=\phi_1]$ is a Gaussian
probability measure centered at (a classical solution
corresponding to) $\phi_1$.

The Hilbert space decompositions (\ref{4.6}) and (\ref{4.7}), and
the isomorphism (\ref{4.8}), imply that a sample field for the
Gaussian $d\hat{\phi}_{C(m_0,\hat{\Sigma })}$ can be uniquely
decomposed as a sum of independent terms:
\begin{equation}\hat{\phi }=\hat{\phi}_0+\hat{\phi}_s=\phi_0+\hat{\phi}_s+\phi_
0^{*},\label{4.11}\end{equation} where $\phi_0$ ($\phi_0^{*}$,
respectively) is a generalized function which is supported on
$\Sigma$ ($\Sigma^{*}$, respectively) and vanishing on $S$, and
$\hat{\phi}_s$ is a solution of the Helmholtz equation in
$\hat{\Sigma}\setminus S$, and determined by its (distributional)
boundary value $\phi_1$ on $S$.  We will write $\phi =\phi_0+\phi_
s$ ($\phi^{*}=\phi^{*}_0+\phi^{*}_s$, respectively) for the
restriction of a random $\hat{\phi}$ to $\Sigma$ ($\Sigma^{*}$,
respectively).

In particular for $a.e.$ $\phi_1$, the $\phi_1$ (normalized)
conditioned measure in (\ref{4.10}) is a direct product
\begin{equation}[d\hat{\phi}_{C(m_0,\hat{\Sigma })}\vert\hat{\phi}\vert_S=
\phi_1]=\end{equation}
\begin{equation}[d\phi_{C(m_0,\Sigma )}\vert\phi_S=\phi_1]\times [d\phi^{*}_{C(
m_0,\Sigma^{*})}\vert\phi^{*}_S=\phi_1]\label{4.12}\end{equation}

\begin{remark} (a) A random $\phi$ for
$[d\phi_{C(m_0,\Sigma )}\vert\phi_S=\phi_1]$ is of the form $\phi
=\phi_0+\phi_s$, $\phi_0$ is a normalized Gaussian with
Cameron-Martin space $W^1_0(m_0,\Sigma )$, and
$(\phi_s)_S=\phi_1$. Thus we could also write
\begin{equation}[d\phi_{C(m_0,\Sigma )}\vert\phi_S=\phi_1]=d\phi_{C(m_0,\Sigma
,D)}(\phi -\phi_s)\label{4.13}\end{equation} where $C(m_0,\Sigma
,D)$ is the inverse of $m_0^2+\Delta_{\Sigma}$ with Dirichlet
boundary condition. The right hand side is defined for all
solutions $\phi_s$ of the Helmholtz equation in $\Sigma\setminus
S$. If one considers a collar $\{0\le t\le\delta \}\times
S^1\subset\Sigma$ for a boundary component ($\{t=0\}$), then the
Collela-Lanford theorem says that for any $\epsilon >0$, with
probability one, $\phi_0$ is a continuous function of $t$ with
values in $W^{-\epsilon}(S^1)$, which vanishes when $t=0$.

(b) To this point we have not given an independent meaning to
$C(m_0,\Sigma )$ or $d\phi_{C(m_0,\Sigma )}$.  The measure
$d\phi_{ C(m_0,\Sigma )}$ can be understood as the Gaussian with
Cameron-Martin space $W^1(\Sigma ,m_0)$ (see (\ref{1.11})); a
random $\phi$ is a restriction of a random $\hat{\phi}$ to
$\Sigma$. However `$C(m_0,\Sigma )=(m_0^2 +\Delta )^{-1}$' does
not have an independent meaning, in reference to $\Sigma$ alone
(because we are interested in a free boundary condition, which is
why we introduce the double of $\Sigma$).
\end{remark}

In terms of this notation, and using reflection symmetry through
$S$, we obtain the following

\begin{lemma}\label{4.14} The pushforward measure $\mathcal Z_1(\Sigma
)^2$ (see Definition \ref{4.3}), is given by
\begin{equation}\left(\int_{\hat{\phi}_0}e^{-\int_{\hat{\Sigma}}:P(\hat{\phi })
:_{C_0}}[d\hat\phi_{C(m_0,\hat{\Sigma })}\vert\hat\phi\vert_
S=\phi_1]\right)d(S_{*}d\hat{\phi}_{C(m_0,\hat{\Sigma })})(\phi_1
)\end{equation}
\begin{equation}=\left(\int e^{-\int_{\Sigma}:P(\phi ):_{C_0}}[d\phi_{C(m_0,\Sigma
)}\vert\phi\vert_S=\phi_1]\right)^2d(S_{*}d\hat{\phi}_{C(m_0
,\hat{\Sigma })})(\phi_1),\end{equation}
\end{lemma}

We now turn to the setup of the theorem.

Suppose that we are given $\Sigma_1$ and $\Sigma_2$. We first
suppose that $\Sigma_1$ has empty incoming boundary, and
$\Sigma_2$ has nonempty outgoing boundary. Thus
$\Sigma_3=\Sigma_2\circ\Sigma_1$ also has empty incoming boundary.

Let $S_1$ denote the outgoing boundary of $\Sigma_1$ (which is the
same as the incoming boundary for $\Sigma_2$), and let $S_2$
denote the outgoing boundary for $\Sigma_2$.  We will write $\phi$
for a field on $\Sigma_1$.  This field has a decomposition $\phi
=\phi_0+\phi_s$, where $\phi_0$ is Gaussian and $\phi_s$ is a
solution of the Helmholtz equation and determined by the boundary
value $\phi_{S_1}$.  We will similarly write $\psi$ for a field on
$\Sigma_2$, with decomposition $\psi =\psi_0+\psi_s$.  We will
also write $\phi_i$ for a field on $S_i$, and $C_i$ will denote
the covariance $(m_0^2+\Delta )^{-1}$ associated to
$\vert\hat{\Sigma}_ i\vert$.

For $\Sigma_3=\Sigma_2\circ\Sigma_1$, there is a finer
decomposition, corresponding to the trace map
\begin{equation}W^1(\Sigma_3)\to W^{1/2}(S_1)\oplus W^{1/2}(S_2)\label{4.15}\end{equation}
and the isomorphism
\begin{equation}W^1_0(\Sigma_3,m_0)=W^1_0(\Sigma_1,m_0)\oplus W^1_0(\Sigma_2,m_
0).\label{4.16}\end{equation} A random field $\Phi$ on $\Sigma_3$
with distribution $d\Phi_{C_3}$ can be written as a sum of
independent Gaussians
\begin{equation}\Phi =\phi_0+\psi_0+\Phi^s,\quad\Phi^s=\phi_s\sqcup\psi_s\label{4.17}\end{equation}
where $\phi_s$ and $\psi_s$ (are random Helmholtz solutions, as
before, and) have common boundary value $\phi_1$ on $S_1$, $\psi_
s$ has boundary value $\phi_2$ on $S_2$, and the
$d\Phi_{C_3}$-distribution for $\Phi^s$, in the coordinates
$(\phi_1,\phi_2)$, is a Gaussian measure with covariance
$(m_0^2+\Delta_{\hat{\Sigma}_3})^{-1}$ restricted to $ S_1\cup
S_2$.  We will write the $d\Phi_{C_3}$ distribution for $\Phi^s$
as $d\Phi^ s_{C_3}$.

\subsection{Part 2. Locality of nonlinear interactions}\label{subs4.2}

We now want to calculate
\begin{equation}\int_{\phi_1\in \mathcal D'(S_1)}\mathcal Z_1(\Sigma_2)\mathcal Z_1(\Sigma_
1)\label{4.18}\end{equation} where the integral is over the common
boundary value $\phi_1=\phi_{S_1}=\psi_{S_1}$.  By Lemma
\ref{4.14} this
\begin{equation}=\int_{\phi_1}\left(\int e^{-\int_{\Sigma_1}:P(\phi ):_{C_0}}[d
\phi_{C_1}\vert\phi_{S_1}=\phi_1])\right)(S_{1*}(d\hat{\phi}_{
C_1}))^{1/2}(\phi_1)\label{4.19}\end{equation}
\begin{equation}\left(\int e^{-\int_{\Sigma_2}:P(\psi ):_{C_0}}[d\psi_{C_2}\vert
\psi_{S_i}=\phi_i)\right)((S_1\sqcup S_2)_{*}(d\hat{\psi}_{C_
2}))^{1/2}(\phi_1,\phi_2)\end{equation}

\begin{proposition}\label{4.20}For a random field $\Phi$ as in
(\ref{4.17}),
\begin{equation}\int_{\Sigma_1}:P(\phi ):_{C_0}+\int_{\Sigma_2}:P(\psi ):_{C_0}
=\int_{\Sigma_2\circ\Sigma_1}:P(\Phi ):_{C_0},\quad a.e.\quad
[d\Phi_{ C_3}]\end{equation}
\end{proposition}

\begin{proof} We first remark that we have not indicated the
dependence of $C_0=-\frac 1{2\pi}log(m_0d(x,y))$ on the underlying
surface, because when there is an ambiguity, the metrics are the
same. The proposition follows from the definition (\ref{3.23}) for
$C_0$-regularization, and Remark 5. \qed
\end{proof}

\begin{corollary}\label{4.21}
\begin{equation}\int_{\phi_1}\mathcal Z_1(\Sigma_2)(\phi_2,\phi_1)\mathcal Z_1(\Sigma_1
)(\phi_1)\label{4.22}\end{equation}
\begin{equation}=\int_{\phi_1}F_P(\phi_2,\phi_1)((S_1\sqcup S_2)_{*}(d\hat{\psi}_{
C_2}))^{1/2}(\phi_1,\phi_2)(S_{1*}(d\hat{\phi}_{C_1}))^{1/2}(\phi_
1)\end{equation} where
\begin{equation}F_P(\phi_2,\phi_1)=\int e^{-\int_{\Sigma_3}:P(\Phi ):_{C_0}}[d\Phi_{
C_3}\vert\Phi_{S_1}=\phi_1,\Phi_{S_2}=\phi_2]\end{equation}
\end{corollary}

\begin{proof} Proposition \ref{4.20}, applied to (\ref{4.19}), implies
that (\ref{4.22}) equals
\begin{equation}=\int_{\phi_1}(\int e^{-\int_{\Sigma_3}:P(\Phi ):_{C_0}}[d\phi_{
C_1}\vert\phi_{S_1}=\phi_1]\times
[d\psi_{C_2}\vert\psi_{S_1}=\phi_
1,\psi_{S_2}=\phi_2])\end{equation}
\begin{equation}((S_1\sqcup S_2)_{*}(d\hat{\psi}_{C_2}))^{1/2}(\phi_2,\phi_1)(S_{
1*}(d\hat{\phi}_{C_1}))^{1/2}(\phi_1)\label{4.23}\end{equation} By
(\ref{4.17}) $d\Phi_{C_3}$ is obtained by (normalized)
conditioning $d\phi_{C_1}\times d\psi_{C_2}$ so that
$(\phi_s)_{S_1}=(\psi_s)_{S_1}$.  Thus
\begin{equation}[d\phi_{C_1}\vert\phi_{S_1}=\phi_1]\times [d\psi_{C_2}\vert\psi_{
S_1}=\phi_1,\psi_{S_2}=\phi_2]\end{equation}
\begin{equation}=[d\Phi_{C_3}\vert\Phi_{S_1}=\phi_1,\Phi_{S_2}=\phi_2]\label{4.24}\end{equation}
Inserting this into (\ref{4.23}) yields the proposition.\qed
\end{proof}

\subsection{Parts 3 and 4. Sewing of normalized background
measures and $\zeta$-regularized volumes}\label{subs4.3}

Now we need to compare the expression in Corollary \ref{4.21} with
$\mathcal Z_1(\Sigma_3)$.  By Lemma \ref{4.14}
\begin{equation}\mathcal Z_1(\Sigma_3)=\int e^{-\int_{\Sigma_3}:P(\Phi ):_{C_0}}[d\Phi_{
C_3}\vert\Phi_{S_2}=\phi_2]((S_2)_{*}(d\hat{\Phi}_{C_3}))^{1/2}(\phi_
2).\end{equation}
\begin{equation}=\int F_P(\phi_2,\phi_1)([d\Phi^s_{C_3}:\Phi^s_{S_2}=\phi_2])((
S_2)_{*}(d\hat{\Phi}_{C_3}))^{1/2}(\phi_2)\label{4.25}\end{equation}
where $F_P$ is as in Corollary \ref{4.21}.

To complete the proof of the Theorem, in comparing (\ref{4.22})
and (\ref{4.25}), it is clear that we need to compare the measures
(with values in half-densities) in the two integrals. These
measures do not depend upon $P$ (all the $P$-dependence is in
$F_P$).

\begin{proposition}\label{4.26} Suppose that $P=0$.  For a.e.
$\phi_2$, the following equality of measures on fields $\phi_ 1$
holds:
\begin{equation}\mathcal Z(\Sigma_2)(\phi_2,\phi_1)\mathcal Z(\Sigma_1)(\phi_1)=\end{equation}
\begin{equation}[d\Phi^s_{C_3}\vert\Phi^s_{S_2}=\phi_2](\phi_1)\mathcal Z(\Sigma_
3)(\phi_2)\end{equation}
\end{proposition}

\begin{remark}\label{4.27}  (a) The measures involved in this
statement are Gaussian, hence eminently computable. The nontrivial
content of the statement involves understanding the way in which
$\zeta$-determinants mesh with the determinants which arise in
calculating compositions of half-densities.

(b) Since $[d\Phi^s_{C_3}\vert\Phi^s_{S_2}=\phi_2]$ is a
probability measure, the free version of the Theorem follows from
this proposition by integrating $1$ on both sides:
\begin{equation}\int_{\phi_1}\mathcal Z(\Sigma_2)(\phi_2,\phi_1)\mathcal Z(\Sigma_1)(\phi_
1)=\mathcal Z(\Sigma_3)(\phi_2).\label{4.28}\end{equation} At the
projective level, this equality has an important interpretation in
terms of the composition of Lagrangian subspaces (see page 147 of
\cite{GS} or \cite{W} for the general definitions).  Given a
$1$-manifold $S$, let $Q(S)$ denote `position space' $W^{1/2}(S)$.
Then as Lagrangian subspaces, the composition of
\begin{equation}W^1_0(\Sigma_1,m)^{\perp}=\{\left(\begin{array}{cc} \phi_1\\
D_{\Sigma_1}\phi_1\end{array} \right)\}\subset
T^{*}Q(S_1)=Q(S_1)\oplus Q(S_1)^{*}\label{4.29}\end{equation} (the
Cameron-Martin space of $\mathcal Z_1(\Sigma_1)^2$, and Helmholtz
solution space on $\Sigma_1$) with
\begin{equation}W^1_0(\Sigma_2,m)^{\perp}=\{\left(\begin{array}{cc} \phi_2\\
A\phi_2+B\phi_1\end{array} \right),\left(\begin{array}{cc} \phi_1\\
-(B^t\phi_2+D\phi_1)\end{array} \right)\}\end{equation}
\begin{equation}\subset T^{*}Q(S_2)\times T^{*}Q(S_1)\label{4.30}\end{equation}
(the Cameron-Martin space of $\mathcal Z_1(\Sigma_2)^2$, and
Helmholtz solution space on $\Sigma_2$, where $D_{\Sigma_2}$ has
been written as a $2\times 2$ matrix, as in (\ref{4.36}) below,
and the minus sign has been inserted because the intrinsic
orientation of $S_1$ is opposite the $\Sigma_2$-induced
orientation (see (\ref{1.18})), is
\begin{equation}W^1_0(\Sigma_2\circ\Sigma_1,m)^{\perp}=\{\left(\begin{array}{cc} \phi_2\\
D_{\Sigma_3}\phi_2\end{array} \right)\}\subset
T^{*}Q(S_2)\label{4.31}\end{equation} (the Cameron-Martin space of
$\mathcal Z_1(\Sigma_3)^2$, and Helmholtz solution space on
$\Sigma_3$).
\end{remark}

\begin{proof} In the course of the proof, we will apply Theorem
B of \cite{BFK} a number of times.  In applying this theorem, when
we consider the Laplacian $\Delta_{\Sigma_i}$, it will be
understood that we are imposing a Dirichlet boundary condition.

Reflecting the decomposition (\ref{4.11}), Theorem B of \cite{BFK}
implies that
\begin{equation}det_{\zeta}(m_0^2+\Delta_{\hat{\Sigma}_i})=det_{\zeta}(m_0^2+\Delta_{
\Sigma_i})det_{\zeta}(2D_{\Sigma_i})det_{\zeta}(m_0^2+\Delta_{\Sigma^{
*}_i})\end{equation}
\begin{equation}=det_{\zeta}(m_0^2+\Delta_{\Sigma_i})^2det_{\zeta}(2D_{\Sigma_i}
).\label{4.32}\end{equation} Reflecting the decomposition
(\ref{4.17}), a slightly extended version of Theorem B implies
that
\begin{equation}det_{\zeta}(m_0^2+\Delta_{\Sigma_3})=det_{\zeta}(m_0^2+\Delta_{
\Sigma_1})det_{\zeta}(m_0^2+\Delta_{\Sigma_2})det_{\zeta}(D_{\Sigma_
1,\Sigma_2})\label{4.33}\end{equation} where
$D_{\Sigma_1,\Sigma_2}$ is the pseudodifferential operator on $
S_1$ which has an inverse with kernel
$(m_0^2+\Delta_{\Sigma_3})^{-1}$.

The statement of the proposition involves half-densities, in the
variable $\phi_2$.  To prove the proposition, it suffices to show
that
\begin{equation}\mathcal Z(\Sigma^{*}_1)(\phi^{*}_1)\mathcal Z(\Sigma^{*}_2)(\phi^{*}_1
,\phi_2)\mathcal Z(\Sigma_2)(\phi_2,\phi_1)\mathcal
Z(\Sigma_1)(\phi_1)=\end{equation}
\begin{equation}[d\Phi^s_{C_3}\vert\Phi^s_{S_2}=\phi_2](\phi_1)\mathcal Z(\Sigma_
3)^2(\phi_2)[d\Phi^{*s}_{C_3}\vert\Phi^{*s}_{S_2^{*}}=\phi_2
](\phi_1^{*}),\label{4.34}\end{equation} as measures on random
fields $\phi_1$, $\phi_2$, and $\phi_1^{*}$.  To clarify the
notation involved in the statement, there is an underlying
factorization
\begin{equation}\hat{\Sigma}_3=\Sigma_1^{*}\circ\Sigma_2^{*}\circ\Sigma_2\circ\Sigma_
1,\label{4.35}\end{equation} and $\phi_1$ is a random field on
$S_1$, the outgoing boundary of $\Sigma_1$, $\phi_2$ is a random
field on $S_2$, the outgoing boundary of $\Sigma_2$, and
$\phi_1^{*}$ is a random field on $S_1^{*}$, the outgoing boundary
of $\Sigma_2^{*}$.  To prove (\ref{4.34}), we will compute the
Fourier transforms of both sides.

Our strategy of proof will involve first doing some intermediate
calculations heuristically (which should serve the dual purpose of
illuminating the meaning of the statements), and then justifying
the answers (by noting that the calculations are valid in finite
dimensions, and taking limits).

We will write $D_{\Sigma_2}$ as a $2\times 2$ matrix,
\begin{equation}D_{\Sigma_2}=\left(\begin{array}{cc} A&B\\
B^t&D\end{array} \right)\label{4.36}\end{equation} relative to the
coordinates $(\phi_2,\phi_1)$.  Thus for example $ D$ has the
following meaning:  given $\phi_1$, calculate the Helmholtz
solution on $int(\Sigma_2)$ which has boundary value $\phi_1$ on
$S_1$ and vanishing boundary value on $S_2$; then $D\phi_ 1$ is
the inward (from the perspective of $\Sigma_2$) normal derivative
along $S_1$.  We will use the two identities:
\begin{equation}D_{\Sigma_1}+D=D_{\Sigma_1,\Sigma_2}\label{4.37}\end{equation}
\begin{equation}A-B(D_{\Sigma_1}+D)^{-1}B^t=D_{\Sigma_3}\label{4.38a}\end{equation}
The first is straightforward.  The second is a coordinate
expression of $(b)$ of Remark \ref{4.27}, because (\ref{4.38a}) is
equivalent to
\begin{equation}D_{\Sigma_3}\phi_2=A\phi_2+B\phi_1,\quad -(B^t\phi_2+D\phi_1)=D_{
\Sigma_1}\phi_1.\label{4.38b}\end{equation}

We will similarly write $D_{\Sigma_2^{*}}$, in terms of $A^{*}$, $
B^{*}$, and $D^{*}$.

In the calculations which follow, we will, in intermediate
heuristic steps, use matrix notation for various pairings. For
example the probability measure $\mathcal Z_1(\Sigma_1)^2$ will be
represented by the heuristic expression
\begin{equation}det(2D_{\Sigma_1})^{1/2}e^{-\frac 12\phi_1^t(2D_{\Sigma_1})\phi_
1}d\phi_1.\label{4.39}\end{equation} We will also use the identity
(valid in finite dimensions)
\begin{equation}det(2D_{\Sigma_2})=det(2D)det(2(A-BD^{-1}B^t)),\label{4.40}\end{equation}
which follows from the factorization
\begin{equation}2D_{\Sigma_2}=\left(\begin{array}{cc} 1&BD^{-1}\\
0&1\end{array} \right)\left(\begin{array}{cc} 2(A-BD^{-1}B^t)&0\\
0&2D\end{array} \right)\left(\begin{array}{cc} 1&0\\
D^{-1}B^t&1\end{array} \right).\end{equation}

We first calculate the Gaussian integral
\begin{equation}\int_{\phi_1}e^{-i(f_1,\phi_1)}\mathcal Z_1(\Sigma_2)(\phi_2,\phi_1
)\mathcal Z_1(\Sigma_1)(\phi_1)=\end{equation}
\begin{equation}\int_{\phi_1}e^{-i(f_1,\phi_1)}det(2D_{\Sigma_2})^{1/4}e^{-\frac
12(\phi^t_2,\phi^t_1)\left(\begin{array}{cc} A&B\\
B^t&D\end{array} \right)\left(\begin{array}{cc} \phi_2\\
\phi_1\end{array} \right)}(d\phi_1d\phi_2)^{1/2}\end{equation}
\begin{equation}det(2D_{\Sigma_1})^{1/4}e^{-\frac 12\phi_1^tD_{\Sigma_1}\phi_1}
(d\phi_1)^{1/2}\end{equation}
\begin{equation}=det(2D_{\Sigma_2})^{1/4}det(2D_{\Sigma_1})^{1/4}\end{equation}
\begin{equation}\int_{\phi_1}exp(-\frac 12(\sqrt {D_{\Sigma_1}+D}\phi_1+\sqrt {
D_{\Sigma_1}+D}^{-1}(B^t\phi_2+if_1))^t\end{equation}
\begin{equation}(\sqrt {D_{\Sigma_1}+D}\phi_1+\sqrt {D_{\Sigma_1}+D}^{-1}(B^t\phi_
2+if_1))d\phi_1\times\end{equation}
\begin{equation}exp(\frac 12(B^t\phi_2+if_1)^t(D_{\Sigma_1}+D)^{-1}(B^t\phi_2+i
f_1))exp(-\frac 12\phi_2^tA\phi_2)(d\phi_2^{})^{1/2}\end{equation}
\begin{equation}=det(2D2D_{\Sigma_1}(D_{\Sigma_1}+D)^{-2})^{1/4}det(2(A-BD^{-1}
B^t))^{1/4}\end{equation}
\begin{equation}e^{-\frac 12f_1^t(D_{\Sigma_1}+D)^{-1}f_1}e^{i\phi_2^tB(D_{\Sigma_
1}+D)^{-1}f_1}\end{equation}
\begin{equation}e^{-\frac 12\phi_2^t\{A-B(D_{\Sigma_1}+D)^{-1}B^t\}\phi_2}(d\phi_
2)^{1/2}\label{4.41}\end{equation} (we also used (\ref{4.40}) in
the last step).

We now calculate, in terms of the identities
(\ref{4.37})-(\ref{4.38a}) (and using reflection symmetry), that
\begin{equation}\int e^{-i((f_1,\phi_1)+(f_2,\phi_2)+(f_1^{*},\phi_1^{*}))}\end{equation}
\begin{equation}\mathcal Z_1(\Sigma^{*}_1)(\phi^{*}_1)\mathcal Z_1(\Sigma^{*}_2)(\phi_1^{
*},\phi_2)\mathcal Z_1(\Sigma_2)(\phi_2,\phi_1)\mathcal
Z_1(\Sigma_1)(\phi_ 1)\label{4.42}\end{equation}
\begin{equation}=det(4D(D_{\Sigma_1}+D)^{-1}D_{\Sigma_1}(D_{\Sigma_1}+D)^{-1})^{
1/2}det(2(A-BD^{-1}B^t))^{1/2}\end{equation}
\begin{equation}\int e^{-\frac 12(f_1^tD_{\Sigma_1,\Sigma_2}^{-1}f_1+f_1^{*t}D^{
-1}_{\Sigma_1^{*},\Sigma_2^{*}}f_1^{*})}e^{i\phi_2^t(BD_{\Sigma_1
,\Sigma_2}^{-1}f_1+B^{*}D^{-1}_{\Sigma_1^{*},\Sigma_2^{*}}f_1^{*}
-f_2)}\end{equation}
\begin{equation}e^{-\frac 12\phi_2^t2D_{\Sigma_3}\phi_2}d\phi_2\end{equation}
\begin{equation}=det(4(1+D_{\Sigma_1}^{-1}D)^{-1}(1+D^{-1}D_{\Sigma_1})^{-1})^{
1/2}det(2(A-BD^{-1}B^t)(2D_{\Sigma_3})^{-1})^{1/2}\end{equation}
\begin{equation}e^{-\frac 12(f_1,(m_0^2+\Delta_{\Sigma_3})^{-1}f_1)+(f_1^{*},(m_
0^2+\Delta_{\Sigma^{*}_3})^{-1}f_1^{*})}\end{equation}
\begin{equation}e^{-\frac 12(B(m_0^2+\Delta_{\Sigma_3})^{-1}f_1+B^{*}(m_0^2+\Delta_{
\Sigma_3^{*}})^{-1}f_1^{*}-f_2),(m_0^2+\Delta_{\hat{\Sigma}_3})^{
-1}(B(m_0^2+\Delta_{\Sigma_3})^{-1}f_1+B^{*}(m_0^2+\Delta_{\Sigma_
3^{*}})^{-1}f_1^{*}-f_2)}\label{4.43}\end{equation}

The expression we have obtained for the Fourier transform is
correct for the following reasons.  Our intermediate calculations
are valid provided that all the objects involved are understood to
be finite dimensional. In particular we can consider compatible
compressions of the positive operators $D_{\Sigma_1}$,
$D_{\Sigma_2}$, $D_{\Sigma_2^{*}}$, and $ D_{\Sigma_1^{*}}$. For
example we can consider the positive operators
$pD_{\Sigma_1}p$, $\left(\begin{array}{cc} p&0\\
0&p\end{array} \right)D_{\Sigma_2}\left(\begin{array}{cc} p&0\\
0&p\end{array} \right)$, $\left(\begin{array}{cc} p&0\\
0&p\end{array} \right)D_{\Sigma^{*}_2}\left(\begin{array}{cc} p&0\\
0&p\end{array} \right)$, and $pD_{\Sigma^{*}_1}p$, where $p$ is
the projection corresponding to a bounded portion of the spectrum
of $D_{\Sigma_1}$ (where $D_{\Sigma_2}$ is written as in
(\ref{4.36}). As the cutoff $p$ is removed, the Gaussian measure
corresponding to $pD_{\Sigma_1}p$ will converge weakly to
$\mathcal Z_1(\Sigma_1)^2$ (the Gaussian corresponding to
$D_{\Sigma_1}$), and so on. We also observe that
\begin{equation}\frac 14(2+D^{-1}D_{\Sigma_1}+D^{-1}_{\Sigma_1}D)\label{4.44}\end{equation}
and
\begin{equation}(A-BD^{-1}B^t)^{-1}(A-B(D_{\Sigma_1}+D)^{-1}B^t)\end{equation}
\begin{equation}=(1-A^{-1}BD^{-1}B^t)^{-1}(1-A^{-1}B(1+D^{-1}D_{\Sigma_1})^{-1}
D^{-1}B^t)\label{4.45}\end{equation} are of the form $1+T$, where
$T$ is trace class.  This is true of (\ref{4.44}), because
$D^{-1}D_{\Sigma_1}=1+H$, where $H$ is Hilbert-Schmidt, hence
\begin{equation}D^{-1}D_{\Sigma_1}+(D^{-1}D_{\Sigma_1})^{-1}=2+T,\quad T=(1+H)^{
-1}H^2,\label{4.46}\end{equation} and $T$ is trace class (the fact
is that $D-D_{\Sigma_1}$ is a smoothing operator, so that $H$
itself is trace class; this follows from use of (\ref{3.22}). This
is true for (\ref{4.45}), because $A^{-1}B$ and $D^{-1}B^t$ are
smoothing operators.  These considerations imply that the
determinants in the last line of (\ref{4.43}) are well-defined.
Furthermore, if we insert the cutoff $p$, the corresponding
determinants will converge, as $p\to 1$. This implies that we can
take a limit of finite dimensional approximations to justify our
formula for the Fourier transform (\ref{4.42}).

We now claim that the Fourier transform of the left hand side of
(\ref{4.34})
\begin{equation}=det(m_0^2+\Delta_{\hat{\Sigma}_3})^{-1/2}e^{-\frac 12(f_1,(m_0^
2+\Delta_{\Sigma_3})^{-1}f_1)+(f_1^{*},(m_0^2+\Delta_{\Sigma^{*}_
3})^{-1}f_1^{*})}\end{equation}
\begin{equation}e^{-\frac 12(B(m_0^2+\Delta_{\Sigma_3})^{-1}f_1+B^{*}(m_0^2+\Delta_{
\Sigma_3^{*}})^{-1}f_1^{*}-f_2),(m_0^2+\Delta_{\hat{\Sigma}_3})^{
-1}(B(m_0^2+\Delta_{\Sigma_3})^{-1}f_1+B^{*}(m_0^2+\Delta_{\Sigma_
3^{*}})^{-1}f_1^{*}-f_2)}\label{4.47}\end{equation} To justify
this claim, we need to show
\begin{equation}det(2D_{\Sigma_1}2D(D_{\Sigma_1}+D)^{-2})^{1/2}det(2(A-BD^{-1}B^
t)(2D_{\Sigma_3})^{-1})^{1/2}\end{equation}
\begin{equation}det_{\zeta}(m_0^2+\Delta_{\hat{\Sigma}_2})^{-1/2}det_{\zeta}(m_
0^2+\Delta_{\hat{\Sigma}_1})^{-1/2}=det_{\zeta}(m_0^2+\Delta_{\hat{
\Sigma}_3})^{-1/2}\label{4.48}\end{equation} Using $(4.32)$ and
$(4.33)$, this is equivalent to
\begin{equation}det(2D_{\Sigma_1}2DD_{\Sigma_1,\Sigma_2}^{-2})^{1/2}det(2(A-BD^{
-1}B^t)(2D_{\Sigma_3})^{-1})^{1/2}\end{equation}
\begin{equation}det_{\zeta}(2D_{\Sigma_1})^{-1/2}det_{\zeta}(2D_{\Sigma_2})^{-1
/2}det_{\zeta}(D_{\Sigma_1,\Sigma_2})det_{\zeta}(2D_{\Sigma_3})^{
1/2}=1\label{4.49}\end{equation} To simplify this, we will use the
well-known fact that $det_{\zeta}(AB)=det_{\zeta}(A)det(B)$, when
$B=1+T$, $T$ trace class (see \cite{Fr} or \cite{KV}). This
implies that (\ref{4.49}) is equivalent to
\begin{equation}det_{\zeta}(2D_{\Sigma_1}2D)^{1/2}det_{\zeta}(2(A-BD^{-1}B^t))^{
1/2}\end{equation}
\begin{equation}det_{\zeta}(2D_{\Sigma_1})^{-1/2}det_{\zeta}(2D_{\Sigma_2})^{-1
/2}=1\label{4.50}\end{equation} Together with the factorization
following (\ref{4.40}), this also implies that
\begin{equation}det_{\zeta}(2D_{\Sigma_2})=det_{\zeta}(2D)det_{\zeta}(2(A-BD^{-
1}B^t)).\end{equation} Thus (\ref{4.50}) is equivalent to showing
that the multiplicative anomaly
\begin{equation}F(2D_{\Sigma_1},2D)=det_{\zeta}(2D_{\Sigma_1}2D)^{1/2}det_{\zeta}
(2D_{\Sigma_1})^{-1/2}det_{\zeta}(2D)^{-1/2}=1\label{4.50}\end{equation}
It is well-known that this vanishes, because $D-D_{\Sigma_1}$ is a
smoothing operator (see \cite{Fr} or \cite{KV}).

We have now established that (\ref{4.47}) is an expression for the
Fourier transform of the left hand side of (\ref{4.34}).

We will now calculate the Fourier transform of the right hand side
of (\ref{4.34}), along the same lines.  As we did for
$D_{\Sigma_2}$, we will write $D_{\Sigma_1,\Sigma_2}$ as a $
2\times 2$ matrix
\begin{equation}D_{\Sigma_1,\Sigma_2}=\left(\begin{array}{cc} \alpha&\beta\\
\beta^t&\delta\end{array} \right)\label{4.51}\end{equation}
relative to the coordinates $(\phi_2,\phi_1)$.  The crucial fact
is that $B=\beta$.

We first calculate (heuristically)
\begin{equation}\int_{\phi_1}e^{-i(f_1,\phi_1)}[d\Phi^s_{C_3}\vert\Phi_{S_2}=\phi_
2]\mathcal Z_1(\Sigma_3)(\phi_2)\end{equation}
\begin{equation}=\int_{\phi_1}e^{-i(f_1,\phi_1)}det(\delta )^{1/2}e^{\frac 12\phi_
2^t(\alpha -\beta\delta^{-1}\beta^t)\phi_2}\end{equation}
\begin{equation}e^{-\frac 12(\phi_2^t,\phi_1^t)\left(\begin{array}{cc} \alpha&\beta\\
\beta^t&\delta\end{array} \right)\left(\begin{array}{cc} \phi_2\\
\phi_1\end{array} \right)}d\phi_1det(2D_{\Sigma_3})^{1/4}e^{-\frac
12\phi_2^tD_{\Sigma_3}\phi_2}(d\phi_2)^{1/2}\end{equation}
\begin{equation}=det(2D_{\Sigma_3})^{1/4}e^{-\frac 12f_1^t\delta^{-1}f_1}e^{i\phi_
2^t\beta\delta^{-1}f_1}e^{-\frac
12\phi_2^tD_{\Sigma_3}\phi_2}(d\phi_
2)^{1/2}\label{4.52}\end{equation}

This implies
\begin{equation}\int e^{-i((f_1,\phi_1)+(f_2,\phi_2)+(f_1^{*},\phi_1^{*}))}\end{equation}
\begin{equation}[d\Phi^s_{C_3}\vert\Phi^s_{S_2}=\phi_2](\phi_1)\mathcal Z_1(\Sigma_
3)^2(\phi_2)[d\Phi^{*s}_{C_3}\vert\Phi^{*s}_{S_2^{*}}=\phi_2
](\phi_1^{*})=\end{equation}
\begin{equation}e^{-\frac 12(f_1,(m_0^2+\Delta_{\Sigma_3})^{-1}f_1)+(f_1^{*},(m_
0^2+\Delta_{\Sigma^{*}_3})^{-1}f_1^{*})}\end{equation}
\begin{equation}e^{-\frac 12(\beta (m_0^2+\Delta_{\Sigma_3})^{-1}f_1-\beta^{*}(
m_0^2+\Delta_{\Sigma_3^{*}})^{-1}f_1^{*}-f_2,(m_0^2+\Delta_{\hat{
\Sigma}_3})^{-1}(\beta (m_0^2+\Delta_{\Sigma_3})^{-1}f_1-\beta^{*}
(m_0^2+\Delta_{\Sigma_3^{*}})^{-1}f_1^{*}-f_2)}\label{4.53}\end{equation}

This last equation is justified, by noting that the calculations
leading to it are valid in finite dimensions and taking limits.

Using $B=\beta$, it is now clear that the Fourier transform of the
right hand side of (\ref{4.34}) equals (\ref{4.47}).  This proves
(\ref{4.34}), and completes the proof of the proposition. \qed
\end{proof}

As we remarked above, this proves part (a) of the Theorem,
assuming that the incoming boundary of $\Sigma_1$ is empty and the
outgoing boundary of $\Sigma_2$ is nonempty. The proofs in the
other two cases for (a) involve straightforward modifications.

To prove (b), suppose that $\Sigma =\Sigma_1\circ\Sigma_2$. Then
\begin{equation}trace(\mathcal Z(\Sigma ))=\mathcal Z(\vert\Sigma_1\vert^{*})\circ \mathcal Z
(\vert\Sigma_2\vert )=\mathcal
Z(\vert\Sigma_1\vert^{*}\circ\vert\Sigma_ 2\vert )=\mathcal
Z(\hat{\Sigma }).\end{equation}
\begin{equation}\label{4.54}\end{equation}
This completes the proof of the Theorem.

\section{Appendix A: Half-Densities}\label{Appendix A}

Suppose that $X$ is a standard Borel space, and $\mathcal C$ is a
measure class on $X$.  Let $\bar {\mathcal C}$ denote the union of
all measure classes which are absolutely continuous with respect
to $\mathcal C$, and let $\bar {\mathcal C}_f$ denote the subset
of finite measures.  There is a real separable Hilbert space,
$\mathcal H(\mathcal C )$, the space of half-densities relative to
$\mathcal C$, and a bilinear map
\begin{equation}\mathcal H(\mathcal C)\times \mathcal H(\mathcal C)\to\bar {\mathcal C}_f,\label{A.1}\end{equation}
which are canonically associated to $\mathcal C$.  We will define
the space of half densities in terms of its representations.

Fix a positive representative $\nu$ for $\mathcal C$.  There is an
isomorphism of Hilbert spaces
\begin{equation}L^2(X,\nu ;\mathbb R)\to \mathcal H(\mathcal C):f\to f(d\nu )^{1/2},\label{A.2}\end{equation}
and in terms of this isomorphism, the map (\ref{A.1}) is given by
\begin{equation}f(d\nu )^{1/2}\otimes g(d\nu )^{1/2}\to fgd\nu\label{A.3}\end{equation}

If one chooses another positive representative for $\mathcal C$,
say $\mu$, then
\begin{equation}f(d\nu )^{1/2}=h(d\mu )^{1/2}\quad\Leftrightarrow\quad f=h(\frac {
d\mu}{d\nu})^{1/2}\label{A.4}\end{equation} where $f\in L^2(d\nu
)$, $h\in L^2(d\mu )$, and $(\frac {d\mu}{d\nu} )^{1/2}$ denotes
the positive square root of this positive function.  In an obvious
way, these identifications can be used to give a formal definition
of $\mathcal H(\mathcal C)$.

We now list a number of elementary facts about spaces of
half-densities.

(1) If $\mathcal C_1<<\mathcal C_2$, then there is a canonical
isometric embedding
\begin{equation}\mathcal H\mathcal C_1)\to \mathcal H(\mathcal C_2):f(d\nu_1)^{1/2}\to f(\frac {
d\nu_1}{d\nu_2})^{1/2}(d\nu_2)^{1/2},\label{A.5}\end{equation}
where $\nu_i$ is a positive representative for $\mathcal C_i$.

(2) The isomorphism (\ref{A.2}), and the coordinate transformation
(\ref{A.5}), show that there is a distinguished positive cone
inside $\mathcal H(\mathcal C)$, corresponding to nonnegative
functions in (\ref{A.2}).  We will denote this cone by $\mathcal
H(\mathcal C )^{+}$. Given a positive finite measure, $\nu$,
$\nu^{1/2}$ will denote the positive square root in the space of
half densities.

(3) The natural representation of $L^{\infty}(\mathcal C)$ by
multiplication operators on $L^2(X,\nu )$ corresponds to a
well-defined natural action
\begin{equation}L^{\infty}(\mathcal C)\times \mathcal H(\mathcal C)\to \mathcal H(\mathcal C):F\otimes
\delta\to F\cdot\delta\label{A.6}\end{equation} Conversely given a
faithful multiplicity free representation of a commutative Von
Neumann algebra
\begin{equation}\mathcal A\times \mathcal H\to \mathcal H,\label{A.7}\end{equation}
there is a measure class $\mathcal C$, unique up to isomorphism,
such that (\ref{A.7}) is realized as (\ref{A.6}).  This is a
special case of the spectral theorem (see \cite{D}, page 210,
theorem 2).

(4) Given disjoint measure spaces $\mathcal C_i$, there is a
canonical isomorphism
\begin{equation}\mathcal H(\mathcal C_i)\oplus \mathcal H(\mathcal C_2)\to \mathcal H(\mathcal C_1\sqcup
\mathcal C_2)\label{A.8}\end{equation}

(5) Given a pair of spaces and measure classes $(X_i,\mathcal
C_i)$, there is a measure class $\mathcal C_1\otimes \mathcal C_2$
on $X_1\times X_2$ generated by $\mathcal C_1\times \mathcal C_2$.
There is a canonical isomorphism
\begin{equation}\mathcal H(\mathcal C_1)\otimes \mathcal H(\mathcal C_2)\to \mathcal H(\mathcal C_1\otimes
\mathcal C_2)\label{A.9}\end{equation}

(6) Suppose that $\nu$ is a finite positive measure belonging the
measure class $\mathcal C\otimes \mathcal C$ on $X\times X$.  Then
\begin{equation}\nu^{1/2}\in \mathcal H(\mathcal C)\otimes \mathcal H(\mathcal C),\label{A.10}\end{equation}
and hence $\nu^{1/2}$ can be interpreted as a Hilbert-Schmidt
operator on $\mathcal H(\mathcal C)$.  This operator is
positivity-preserving:
\begin{equation}\nu^{1/2}:\mathcal H(\mathcal C)^{+}\to \mathcal H(\mathcal C)^{+}\label{A.11}\end{equation}
(in \cite{Simon}, page 30, the phrase `doubly Markovian map' is
used for this property).

Given $\nu_1$ and $\nu_2$,
\begin{equation}\nu_1^{1/2}\circ\nu_2^{1/2}=\nu_3^{1/2},\label{A.12}\end{equation}
where $\nu_3$ is another finite positive measure,
\begin{equation}\nu_3(\phi ,\psi )=(\int_{\eta}\nu_1(\phi ,\eta )^{1/2}\nu_2(\eta
,\psi )^{1/2})^2.\label{A.13}\end{equation} This can be summarized
as follows.

\begin{proposition}\label{A.14}The finite positive measures in
$\mathcal C\otimes \mathcal C$ form a semigroup, with
multiplication (\ref{A.12}), and this semigroup is represented by
positivity-preserving Hilbert-Schmidt operators on $\mathcal
H(\mathcal C)$.
\end{proposition}

(7) Now suppose that the Borel structure on $X$ is derived from a
locally convex linear structure, and $\mathcal C$ is the measure
class on $X$ of a Gaussian measure (chapter 2 of \cite{B}).

\begin{proposition}\label{A.15} Suppose that $\nu_1$ and $\nu_2$
are Gaussian measures belonging to $\mathcal C\otimes \mathcal C$.
Then $\nu_ 3$, as in (\ref{A.12})-(\ref{A.13}), is a multiple of a
Gaussian measure.
\end{proposition}

If $X$ is finite dimensional, this is a consequence of the Theorem
in section 3, page 65 of \cite{Howe}.  The Proposition follows in
a routine way, after rewriting Howe's formulas to account for
normalizations of measures, by taking limits.

\begin{acknowledgements} I thank Lennie Friedlander and
John Palmer for useful conversations. I also thank a referee for
pointing out a serious error in a first version of this paper; see
Remark \ref{3.25}.
\end{acknowledgements}

\end{document}